\documentstyle[prb,aps]{revtex}
\input epsf
\input psfig.sty 

\begin{document}
\bibliographystyle{prsty}
\title{Effect of small-scale architecture on polymer mobility} 

\author{Jutta Luettmer-Strathmann}
\address{
Department of Physics, The University of Akron, Akron, OH 44325-4001}
\date{\today}
\maketitle
\begin{abstract}
Processes on different length scales affect the dynamics of chain molecules.
In this work, we focus on structures on the scale of a monomer and 
investigate polyolefins, i.e.\ hydrocarbon chains with different small scale
architectures.
We present an exact enumeration scheme for the simulation of 
interactions and relative motion of two short chain sections 
on a lattice and employ it 
to deduce the probability for segmental motion for 
polymers of four different architectures in the melt.
The probability for segmental motion is inversely proportional to the 
monomeric friction coefficient and hence the viscosity of a polymer.
Combining our simulation results with an equation of state for the
thermodynamic properties of the polymers, we are able to make predictions
about the variation of the friction coefficient 
with temperature, pressure, and small scale architecture. 
To compare our results with experimental data, 
we have determined monomeric friction coefficients from experimental
viscosity data for the four polyolefins considered in this work.
For temperatures well above the glass transition temperature, we 
find that our simple approach gives a good qualitative representation
of the variation of the friction coefficient with chain architecture, 
temperature and pressure. 
\end{abstract}

\section{Introduction}\label{intro}

\begin{figure}
\epsfxsize=13cm
\epsfbox{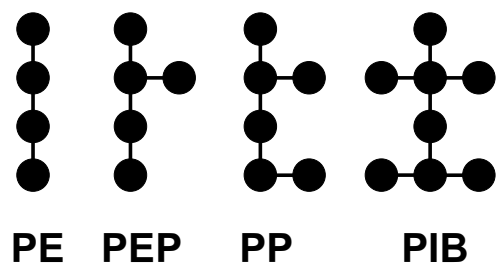}
\vspace{1cm}
\caption{United atom representation of 
the polyolefins considered in this work. Shown are the repeat units
with four carbon atoms in the backbone for polyethylene (PE), 
an alternating copolymer of polyethylene and polypropylene (PEP), 
polypropylene (PP), and 
polyisobutylene (PIB).}
\label{fig1}
\end{figure}

The dynamics of chain molecules are affected by 
local interactions between individual chain segments 
as well as processes on the length scale of the whole chain
and collective motions of chain segments.
In this work, we investigate the effect of small-scale  
chain architecture on the dynamic properties of polymers.
The polyolefins depicted in 
Fig.~\ref{fig1} are a good example for this effect.  
These polyolefins, all hydrocarbons with sum formula
C$_n$H$_{2n}$, differ considerably in their viscoelastic 
properties~\cite{be68,fe94,ng96,pe87,pe94,ge97,pe88,fe91,ba84b,vo91}
despite their chemical similarity.

An important ingredient in theories for polymer dynamics is a 
friction coefficient $\zeta$ 
which is employed in coarse-grained models to describe small-scale
effects on the dynamics of the system (cf.\ Ref.~\onlinecite{do86}). 
In the Rouse model, for example, chain segments consisting of many
monomers are represented by a single bead and spring.  
The Rouse friction coefficient $\zeta_{\text R}$ describes the 
damping of the bead motion by the surrounding medium.
The Rouse viscosity $\eta_{\text R}$ is proportional to $\zeta_{\text{R}}$ 
and given by~\cite{do86}
\begin{equation}\label{eq1}
\eta_{\text{R}}=\frac{N_A}{36}\rho\frac{R_0^2}{M}N_R\zeta_{\text{R}},
\end{equation}
where $N_A$ is Avogadro's number, $\rho$ is the mass density, 
$M$ is the molecular mass, 
$R_0^2$ is the mean-squared end-to-end distance of the chain, 
and $N_R$ is the number of Rouse segments per chain.
The Rouse model describes the viscosity of polymer melts of sufficiently short 
chains, i.e.\ for polymers with a molecular mass well below the 
entanglement mass $M_{\rm e}$. For larger molecular masses, entanglement
effects have to be taken into account.\cite{do86}
While $\rho$ and $R_0^2$ vary with temperature at constant pressure, 
almost all of the temperature dependence of the viscosity 
is contained in the friction coefficient.\cite{be68} 
Since the viscosity increases strongly as the glass transition 
temperature is approached, the friction coefficient $\zeta$ reflects 
collective motion as well as small-scale interactions.

The friction coefficient $\zeta$ is inversely proportional to the 
probability for segmental motion,\cite{bu52} a relationship 
that we are going to
exploit in this work.
In an exact enumeration procedure we perform lattice simulations of
relative motion and interactions of two short chain segments, where
the surrounding medium is represented in an average way. 
The collected statistics are evaluated as described in 
Section~\ref{sim} to yield the 
the average probability for segmental motion as a function of 
a reduced temperature and lattice filling fraction 
for each of the architectures in Fig.~\ref{fig1}.

In Section~\ref{zeta} we combine our simulation results with the 
recently developed Born-Green-Yvon lattice model for the
thermodynamic properties of polymers~\cite{li98,li92b,lu99} 
to investigate the mobility of the four polyolefins at given 
temperature and pressure. 
Our method of determining the friction coefficients is not an
absolute one, but 
using the probability for segmental motion of a linear chain 
(polyethylene in our case) at 413~K and atmospheric pressure
as a reference value we are able to calculate 
relative values of the 
friction coefficients as a function of temperature and pressure. 

In order to compare our results with experimental data, 
we extract friction coefficients of the polyolefins of interest
from experimental viscosity data as described in Section~\ref{exp}.
For temperatures well above the glass transition temperature, we 
find that our approach gives a good qualitative representation
of the variation of the friction coefficient with temperature and
chain architecture. We also compare the pressure dependence of
the viscosity of polypropylene~\cite{ma97e} with our predictions and find
very good agreement for low to moderate pressures.
The results presented here as well as future 
directions of this work are discussed in Section~\ref{disc}.

\section{Simulation of local mobility}\label{sim}

A central point of this work is the determination of the probability
of segmental motion from a consideration of two short chain segments
in a dense medium. To this end we perform an exact enumeration of
all possible combined configurations and relative movements of two chain
segments on a lattice. During the enumeration procedure we collect statistics
on the characteristic parameters of each possible initial and final
configuration and the connecting move. In a second step, these
statistics are evaluated for conditions corresponding to different
temperatures and densities. The advantage of this two-step procedure
is that the time-consuming part, the exact enumerations, have to be performed
only once to yield results that can be evaluated quickly for a
variety of conditions.

\subsection{Procedure}\label{exact}

\begin{figure}
\epsfxsize=13cm 
\epsfbox{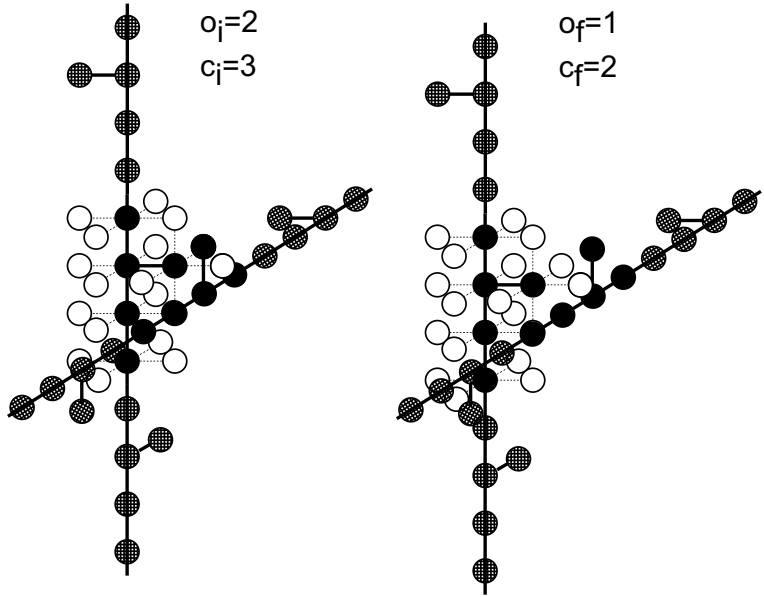}
\vspace{1cm}
\caption{Illustration of the simulation procedure. The figure
on the left shows an initial combined configuration of two 
PEP segments. There are $s_f=5$ sites in the section of interest
in each chain, indicated by the dark filled circles, and $n_{\rm t}=18$ 
identified nearest neighbor sites of chain one (indicated by open circles). 
The numbers $o_i$ and $c_i$ indicate the number of nn sites occupied
by chain two and the number of established contacts, 
respectively.
The figure on the right shows the new combined configuration after
chain one has been moved one lattice site to the right.
The numbers $o_k$ and $c_k$ give occupied sites and established
contacts, respectively. In the move,  $s_{\rm n}=5$ sites
were newly occupied by the first segment.}
\label{fig2}
\end{figure}

For each of two polymer molecules, referred to as chain one and 
chain two from now on, we consider a straight
section composed of three repeat units with given (generally not identical)
side group arrangements. The repeat unit in the
middle is the section of interest in each case, while the attached units
represent the rest of the (long) chains.
As depicted in Fig.~\ref{fig1}, we employ repeat units with four carbon
atoms in the backbone for all the polyolefins considered here.
The simulation procedure is illustrated in Fig.~\ref{fig2}.

In a preparatory step, all possible conformations (side-group
arrangements of the three-repeat-unit sections) of the chains 
are generated. The basic step in the enumeration, to be described below, 
is performed for given conformations and relative orientation of the chains.  
Keeping the conformation and orientation of chain one fixed, 
the basic step is repeated for all orientations of chain two. 
This sequence of steps is then repeated for different chain-two 
conformations until all conformations of chain two are exhausted. 
Then the whole sequence is repeated for all different chain-one 
conformations. This assures that we enumerate the results for all 
possible and distinct combined configurations of the chains.

In a basic step,  
the section of interest of chain one is fixed to the origin of
a simple cubic lattice and aligned with the $z$-axis.
The total number $n_{\rm t}$ and the coordinates of the  non-bonded nearest 
neighbors (nn) sites are determined as is the maximum number 
$c_{\rm m}=4s_{\rm f}$ of possible contacts, where
$s_{\rm f}$ denotes the number of lattice sites occupied by the  
section of interest. For a polyethylene (PE) chain, for example, the 
section of interest occupies $s_{\rm f}=4$ sites,  
has $n_{\rm t}=16$ nearest neighbor sites and a maximum of 
$c_{\rm m}=16$ contacts, while for polyisobutylene (PIB) 
the values are $s_{\rm f}=8$, $n_{\rm t}=24$, and $c_{\rm m}=32$.

Next, an orientation for chain two is chosen, and the following procedure
is repeated for each site of interest of chain two and each identified
nearest neighbor site of chain one, where care has to be taken to avoid double
counting. For the given orientation of chain two, contact is made between
the chains by moving the currently considered site of chain two 
onto the currently considered nearest neighbor site of chain one. 
If the sections do not overlap, the combined configuration is accepted.
It is evaluated by counting how many of the $n_{\rm t}$ nearest neighbor 
sites of chain one are occupied by chain two (this number is called $o_i$) 
and by counting the number $c_i$ 
of contacts between chains established in this way. 
The numbers $(o_i,c_i)$ characterize the static
properties of the initial state. 

In order to determine the mobility of the 
segments, an attempt is made to displace chain one by one lattice
site in each of the six directions, $\pm x$, $\pm y$, and $\pm z$, in turn. 
If the attempt leads to overlap between the chains, it is counted as 
impossible. Otherwise, the number $s_{\rm n}$ of lattice sites newly 
occupied by
the first section of interest is counted and the new combined configuration 
is characterized by determining the numbers $o_f$ and $c_f$ of 
occupied nn sites and established contacts, respectively.
The numbers $(o_f,c_f)$ characterize the static
properties of the final state, while the set 
$(i,f,s_{\rm n})\equiv
(o_i,c_i;o_f,c_f;s_{\rm n})$
characterizes the move.
 
For the results presented here, the exact enumeration procedure described
above was performed for PE, PEP, and PP. 
To avoid excessive computation times for PIB, we generated 
representative samples of one eighth of the single chain conformations 
and proceeded with those as described above.
The different representative samples for the PIB conformations give 
essentially identical results for the properties presented in this work.
The result of the simulations are the frequency of occurrence, 
$n(i,f,s_{\rm n})$, of moves of type $(i,f,s_{\rm n})$, 
as well as statistics on the type, $(o_k,c_k)$, and 
frequency of occurrence, $m_k$, of the combined configurations.

\subsection{Evaluation}\label{eval}

In order to determine the probability $P$ for segmental motion, 
we consider the probabilities $P(i,f,s_{\rm n})$ for the different
types of moves and then form the sum:
\begin{equation}\label{eval1}
P=\sum_{(i,f,s_{\rm n})} P(i,f,s_{\rm n}).
\end{equation}
 The probability $P(i,f,s_{\rm n})$ for a move of type 
$(i,f,s_{\rm n})$ is expressed as 
\begin{equation}\label{eval2}
P(i,f,s_{\rm n})=
\frac{n(i,f,s_{\rm n})P_iP_{\Delta E}P_{\phi}}
{\displaystyle \sum_{f,s_{\rm n}} n(i,f,s_{\rm n})}, 
\end{equation}
where $P_i$ is the probability for the initial combined configuration
to be of type $(o_i,c_i)$, 
$P_{\Delta E}$ accounts for the energy difference $\Delta E=E_f-E_i$ 
between initial and final states, 
and $P_{\phi}$ is the probability that a sufficient number of contiguous sites 
is available to the moving segment in a lattice filled to a fraction $\phi$.
 
Both $P_i$ and $P_{\Delta E}$ involve the energy of a combined configuration
of chain segments in a dense medium. As explained in detail in 
Refs.~\onlinecite{lu99} and \onlinecite{lu99b}, 
the energy $E_k$ for a combined 
configuration characterized by $(o_k,c_k)$ is obtained from 
\begin{equation}\label{eval3}
E_k=\epsilon
\left(c_k+\frac{(n_{\rm t}\xi-o_k)(c_{\rm m}-c_k)}
{n_{\rm t}-o_k} \right),
\end{equation}
where $\epsilon$ is the interaction energy between two molecular sites 
($\epsilon<0$) and where $\xi=2\phi/(3-\phi)$ is the contact density
for infinitely long chains at a filling fraction of $\phi$.
The contributions to the energy $E_k$ in Eq.~(\ref{eval3})
are due to contacts between chain one and chain two and 
between chain one and its randomly filled nearest 
neighbor sites, respectively. 
Please note that in the case of unbranched segments (PE), 
Eq.~(\ref{eval3}) implies $E_k=\epsilon c_{\rm m}\xi$ 
for all $k$ so that there are no energetically preferred configurations 
for PE. The unbranched segments thus serve as our reference system
which allows the isolation of the effects of small scale architecture.

The probability $P_i$ for an initial combined configuration 
with energy $E_i$ is proportional to the Boltzmann factor:
\begin{equation}\label{eval4}
P_i=m_i e^{-\beta E_i} / \sum_{k} m_k e^{-\beta E_k},
\end{equation}
where $m_i$ is the multiplicity of the combination $(o_i,c_i)$
and where $\beta=1/k_{\rm B}T$ with temperature $T$ 
and Boltzmann's constant $k_{\rm B}$.
The effect of energetics on the probability of a move is described 
using the Metropolis form
\begin{equation}\label{eval5}
P_{\Delta E}=\left\{\begin{array}{ll}
e^{-\beta\Delta E}  & \mbox{ if $\Delta E>0$} \\ 
1  & \mbox{ if $\Delta E\leq 0$} 
\end{array} , \right.
\end{equation}
where $\Delta E=E_f-E_i$.

The mobility of a chain segment  is greatly reduced by the presence of 
the other chains in its surroundings. Consider, for the moment, monomers
on a lattice and the attempt of a single particle 
to move from one site to a neighboring site. If all of its neighboring
sites are occupied, the attempt will certainly fail. But even if a nearest
neighbor site is available, the attempt may fail when another particle
is headed for the same site. The only way to guarantee that an attempted
move will be successful is to require that a neighboring site as well as its nearest neighbor
sites (except for the one occupied by the particle under consideration)
are empty. This is a total of six sites for monomers on a simple cubic lattice.
Extending this reasoning to chain molecules, where each monomer is 
bonded (on average) to two monomers on neighboring sites, we require 
four empty sites in the neighborhood of each monomer involved in the move.
In terms of our variables introduced above, a volume of $4s_{\rm n}$ is
required in a move in which $s_{\rm n}$ sites are newly occupied. 
Assuming a random distribution of voids over the lattice, the 
probability of finding $4s_{\rm n}$ lattice sites among the $n_{\rm t}$
nearest neighbor sites of the segment of interest is given by \cite{co59}
\begin{equation}\label{eval6}
P_{\phi}=\exp{\left(-\frac{4s_{\rm n}}{n_{\rm t}(1-\phi)}\right)},
\end{equation}
where $(1-\phi)$ is the fraction of empty sites. 

In Fig.~\ref{fig3} we present simulation results for the probability 
of segmental motion, evaluated according to Eq.~(\ref{eval1}) 
with Eqs.~(\ref{eval2}) to (\ref{eval6}). The probability $P$ is shown 
as a function of reduced temperature $T/T^*=k_{\rm B}T/\epsilon$ 
for a given filling fraction $\phi$ for the four architectures considered in
this work. The effect of the increasing number of side groups on the
mobility is clearly visible. The linear chain (PE) has the highest 
probability of segmental motion followed by PEP, PP, and PIB which have
one, two, and four side groups in the four-carbon backbone monomer, 
respectively. The insert shows the probability $P$ as a function of 
reduced temperature for three different filling fractions 
for the PP architecture. As expected, the probability for segmental motion 
increases with reduced temperature and decreases with filling 
fraction.

\section{Calculation of friction coefficients}\label{zeta}

With a method to obtain the probability for segmental motion in hand, 
we are now in a position to address the monomeric friction coefficient 
$\zeta\propto P^{-1}$. The proportionality constant between $P^{-1}$ and
$\zeta$ is not easily determined. However, 
as pointed out earlier, our goal is to determine 
how the local architecture changes the mobility of branched chains compared to 
that of linear chains.
Hence, we choose a reference state for the linear chain and express our
results for the friction coefficients as the ratio
\begin{equation}\label{zeta1}
\frac{\zeta}{\zeta_{\text{ref}}}=\frac{P_{\text{ref}}}{P}, 
\end{equation}
where $P_{\text{ref}}$ and $\zeta_{\text{ref}}$ are the reference state 
values of the 
probability of segmental motion and the friction coefficient
of the linear chain, respectively.
The reference state can be chosen freely; in our case a temperature of
$T_{\text{ref}}$=413.15~K and a pressure of $p_{\text{ref}}$=0.1~MPa 
turn out to be convenient.
In order to make contact with experimental data we employ equations 
of state based on the recently developed Born-Green-Yvon (BGY) lattice 
model.\cite{li98,li92b} The BGY lattice model has 
three system-dependent parameters for a polymer melt, corresponding
to the volume $v$ per lattice site, the number $r$ of sites occupied by each
chain, and the interaction energy $\epsilon$ between non-bonded nearest 
neighbors. For each of the polymers considered in this work, 
values for the system-dependent parameters have been determined from a 
comparison with experimental temperature-density-pressure data~\cite{lu99} 
and are summarized in Table~\ref{tab1}. 

\begin{figure}
\psfig{figure=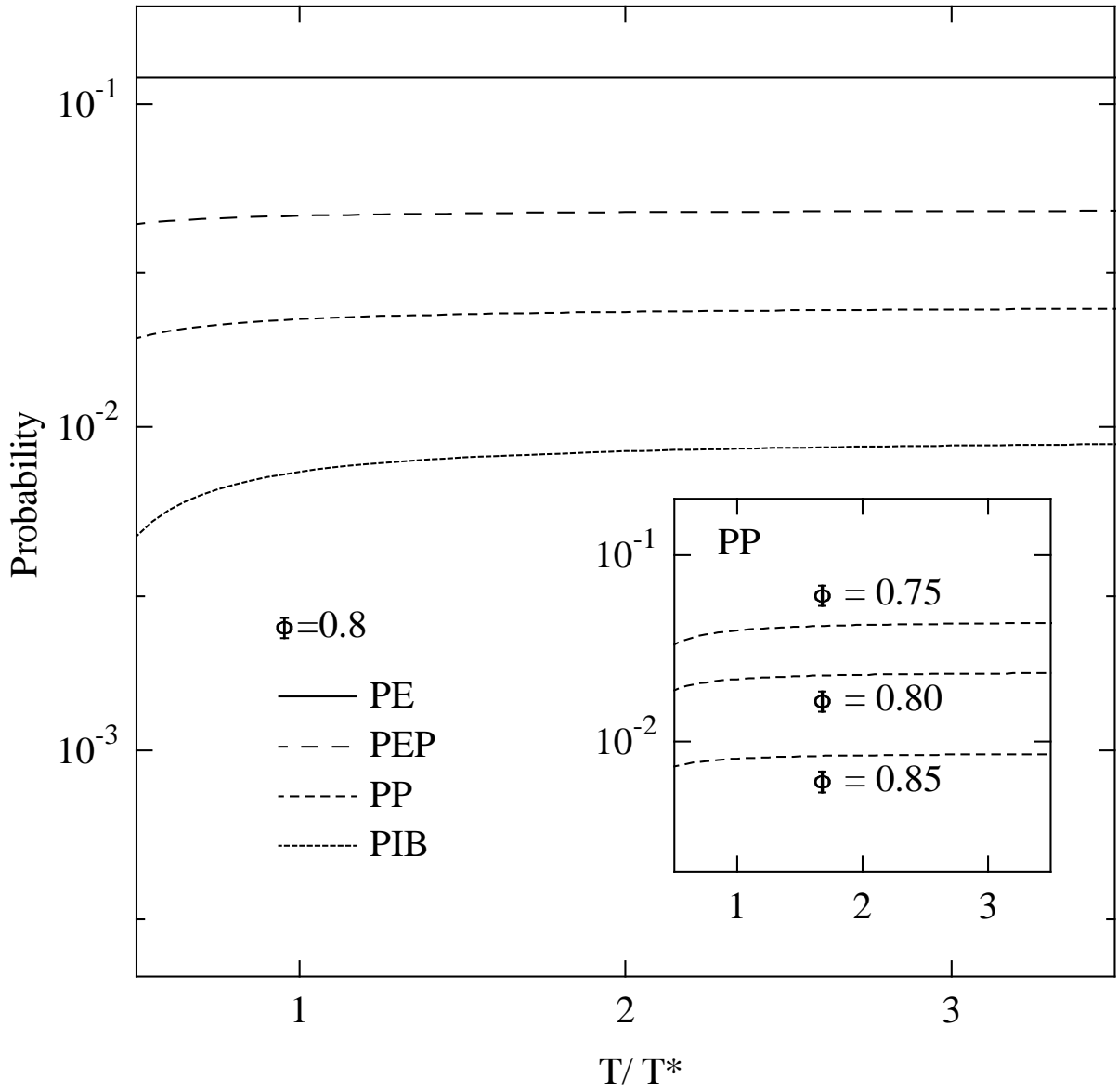,width=16cm}
\vspace{1cm}
\caption{Probability for segmental motion as 
a function of reduced temperature $T/T^*=k_{\rm B}T/\epsilon$ 
at constant filling fractions $\phi$ 
determined from the simulation procedure and 
Eqs.~(\protect{\ref{eval1}}) -- (\protect{\ref{eval6}}).
The graph shows results for the four different architectures 
(see Fig.~\protect{\ref{fig1}}) 
at a common filling fraction of $\phi=0.8$, while the 
insert displays results for three different filling fractions for 
the architecture of PP.
}
\label{fig3}
\end{figure}

\begin{figure}
\psfig{figure=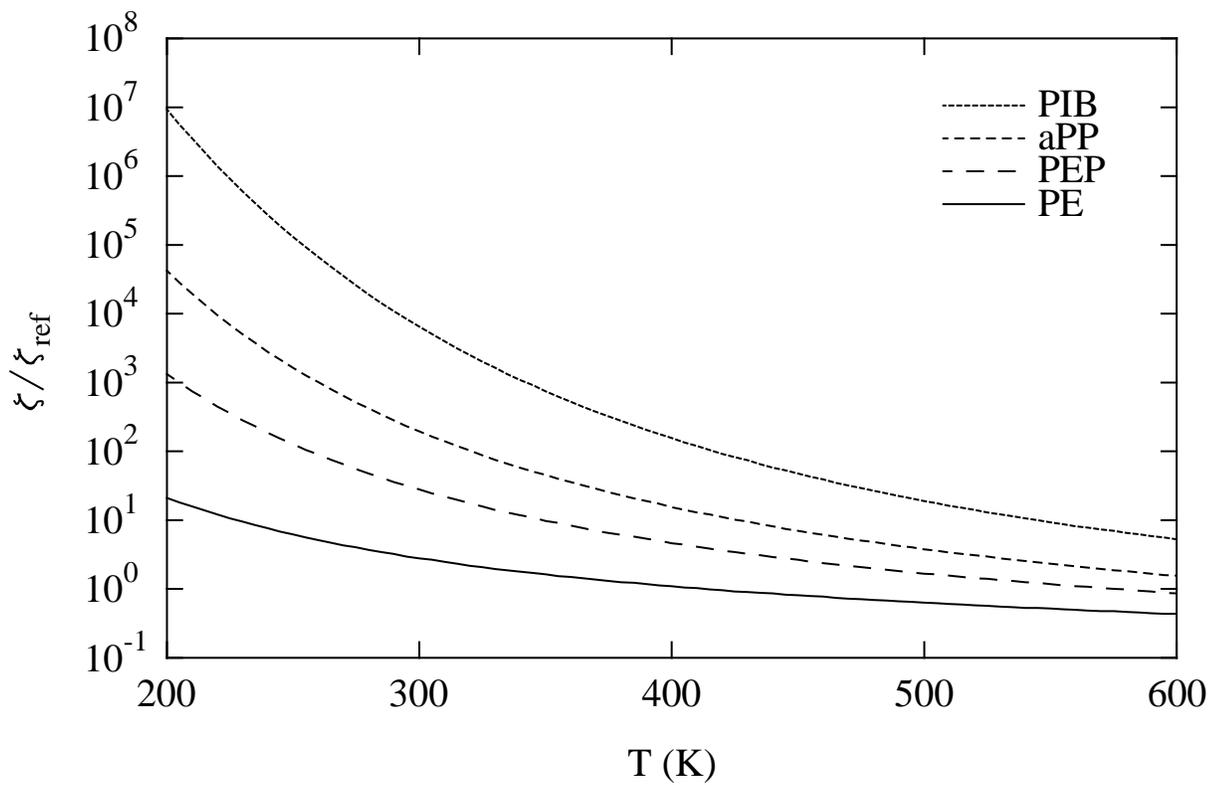,width=16cm}
\vspace{1cm}
\caption{Calculated friction coefficients $\zeta/\zeta_{\text{ref}}$, 
as a function of temperature at constant pressure $p=0.1$MPa for the 
polyolefins considered in this work. The reference value is obtained
for the linear chain (PE) at 413.15~K and 0.1~MPa.} 
\label{fig4}
\end{figure}

In Fig.~\ref{fig4} we present calculated values for the relative 
friction coefficient $\zeta/\zeta_{\text{ref}}$ as a function of temperature 
at a pressure of 0.1~MPa for the polyolefins considered in this work.
Please note that the temperature variation here is much larger than
that in Fig.~\ref{fig3}. While Fig.~\ref{fig3} depicts
the probability $P\propto \zeta^{-1}$ at constant $\phi$, i.e.\ at
constant density $\rho=\phi/rv$,  Fig.~\ref{fig4} shows constant
pressure results. It is the temperature variation of the density at
constant pressure that, through Eq.~(\ref{eval6}), is responsible for 
the strong temperature dependence of $\zeta/\zeta_{\text{ref}}$.

\section{Comparison with experimental data}\label{exp}

The friction coefficient $\zeta$ is not a directly measured quantity
but can be extracted from measurements of dynamic properties like
the viscosity or the self-diffusion coefficient. 
The most direct access to $\zeta$ is through the Rouse viscosity 
$\eta_{\rm R}$ described in the Introduction. 
With the aid of Eq.~(\ref{eq1}) a friction coefficient per monomer
can be defined as follows:
\begin{equation}\label{exp1} 
\zeta=\frac{N_{\rm R}\zeta_{\rm R}}{N}=m_0\frac{\eta_{\text{R}}}{M}
\left(\frac{N_A}{36}\rho\frac{R_0^2}{M}\right)^{-1}, 
\end{equation}
where $N$ is the degree of polymerization and $m_0$ is the mass of a 
monomer which we take to be $n\times 14.03$ g/mol, where $n$ is the number
of carbon atoms in the repeat unit depicted in Fig.~\ref{fig1}.
In addition to the Rouse viscosities $\eta_{\rm R}/M$, 
evaluation of Eq.~(\ref{exp1}) requires values for 
$R_0^2$, the mean-squared end-to-end distance of the chains, 
and for $\rho$, the mass density of the melt. 
In this work, we employ experimental values for these properties
at a temperature of 413~K presented in a recent review by 
Fetters {\it et al.}~\cite{fe94} and included in Table~\ref{tab2}.

\begin{figure}
\psfig{figure=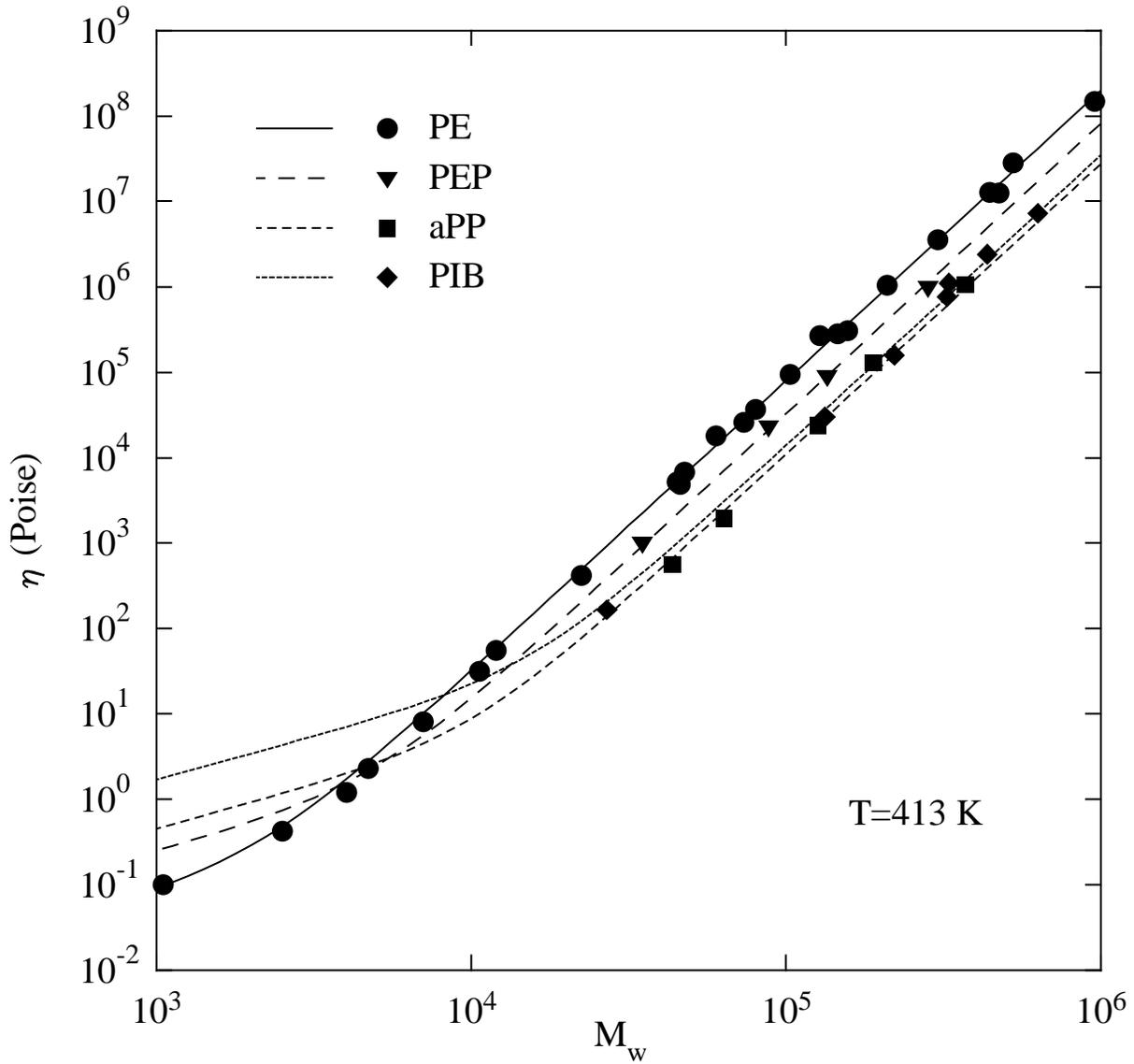,width=16cm}
\vspace{1cm}
\caption{Viscosity of polyolefin melts as a function of molecular mass 
$M_{\rm w}$. The symbols represent experimental data adjusted to 413~K  
as explained in the text. The references for the data are 
Refs. \protect\onlinecite{pe94,ba84b,vo91} for PE, 
\protect\onlinecite{ge97} for PEP,
\protect\onlinecite{pe88} for aPP, and  
\protect\onlinecite{fe91} for PIB. 
The lines represent the empirical scaling law Eq.~(\protect\ref{exp2}), 
where $M_{\rm c}=2.2M_{\rm e}$ with
$M_{\rm e}$ from Ref. \protect\onlinecite{fe94},
and where the values for $\eta_{\rm R}/M$ have been determined 
from a comparison with the experimental data.  
Values for these system-dependent
parameters are included in Table~\protect\ref{tab2}.}
\label{fig5}
\end{figure}

The Rouse model describes directly the viscosity for melts of low molecular
mass $M\ll M_{\rm e}$ (for example, $M_{\rm e}\approx 1000$ for PE and 
$M_{\rm e}\approx  7300$ for PIB~\cite{fe94}). Unfortunately, 
measurements of the 
melt viscosities for short chains are not only scarce, but chain-end
effects may have to be taken into account in their evaluation.\cite{pe87}  
We therefore decided to turn to high molecular-weight viscosity data and
an empirical scaling relation~\cite{gr81} to extract values for 
$\eta_{\rm R}/M$.
Motivated by the reptation model~\cite{do86} and by experience with 
experimental data, Graessley and Edwards~\cite{gr81} suggested the 
following molecular mass dependence of viscosities in polymer melts:
\begin{equation}\label{exp2}
\eta=
\frac{\eta_{\rm R}}{M}\; 
M\left[1+\left(\frac{M}{M_{\rm c}}\right)^{2.4}\right],
\end{equation}
with $M_{\rm c}=2.2M_{\rm e}$.\cite{gr81,oc92}
In principle, Eq.~(\ref{exp2}) could be compared to experimental 
viscosity data at a given temperature to extract both 
$\eta_{\rm R}$ and $M_{\rm c}$. 
This, however, is not advisable since the results for $\eta_{\rm R}$ 
and $M_{\rm c}$ will then strongly depend on the range of molecular
weights for which viscosity data are available. 
Instead, we fix $M_c=2.2M_{\rm e}$, taking the experimental 
$M_{\rm e}$ values at 413~K of Fetters {\it et al.}~\cite{fe94}
quoted in Table~\ref{tab2}, and fit for $\eta_{\rm R}/M$ only.  
This procedure requires values for the viscosity at 413~K, which
we obtain by employing the temperature correlations provided with the 
experimental data~\cite{pe94,ge97,pe88,fe91} to shift the
viscosity values from the temperature of the measurements to 413~K.
More information on the temperature correlations is provided below.
In Fig.~\ref{fig5} we present the data (symbols) 
for the viscosity at 413~K obtained
in this way as a function of molecular mass for the polyolefins considered 
in this work. 
The resulting correlations for the mass
dependence of the viscosity are represented by the lines in 
Fig.~\ref{fig5} and are seen to give a satisfactory representation of
the experimental data. The values for $\eta_{\rm R}/M$ obtained in this way
are included in Table~\ref{tab2}. Inserting them 
into Eq.~(\ref{exp1}) and employing the values for 
$m_0$, $R_0^2$, and $\rho$ as discussed, we arrive at the values
for the monomeric friction coefficient $\zeta$ at 413~K and 
atmospheric pressure presented in Table~\ref{tab2}.
The value of the friction coefficient for PE at 413~K and 0.1~MPa
is the reference value for the experimental friction coefficients,
and all further results will be presented as $\zeta/\zeta_{\rm PE}$(413~K).

In measurements of viscoelastic properties of polymers, it is 
customary to describe the temperature dependence of the viscosity
by equations of the type~\cite{be68} Vogel-Tammann-Fulcher (VTF) 
\begin{equation}\label{exp3}
\ln(\eta(T))=\ln(A)+\frac{1}{\alpha(T-T_0)} ,
\end{equation}
or Williams-Landel-Ferry (WLF) 
\begin{equation}\label{exp4}
\log(\eta(T))=\log(\eta(T_s))-\frac{C_1(T-T_s)}{C_2+T-T_s} . 
\end{equation}
These equations are equivalent with constants related by 
$C_2=T_s-T_0$, $C_1C_2\ln(10)=1/\alpha$, and 
$\ln(A)=\ln(\eta(T_s))-C_1\ln(10)$. Here $T_s$ is an arbitrary reference
temperature while $T_0$ indicates the temperature where the system
is no longer able to relax to an equilibrium state in a finite
amount of time. 
Since the dominant contribution to the temperature variation of 
the viscosity is due to the friction coefficient,
it is a reasonable approximation to assign the temperature 
dependence of the viscosity to the friction coefficient.\cite{be68}
Some of the experimental works quoted here provide slightly different
temperature correlations for the viscosity of samples of different 
molar masses. When shifting the experimental viscosity data to the 
reference temperature of 413~K, we employed the correlations
appropriate for the molar mass under consideration. 
For the following comparison with our work, on the other hand, we choose 
a representative correlation for each polyolefin and bring it into
WLF form with a reference temperature of $T_{\text{ref}}$=413~K. The
corresponding parameters $C_1$ and $C_2$ are included in Table~\ref{tab2}.

\begin{figure}
\psfig{figure=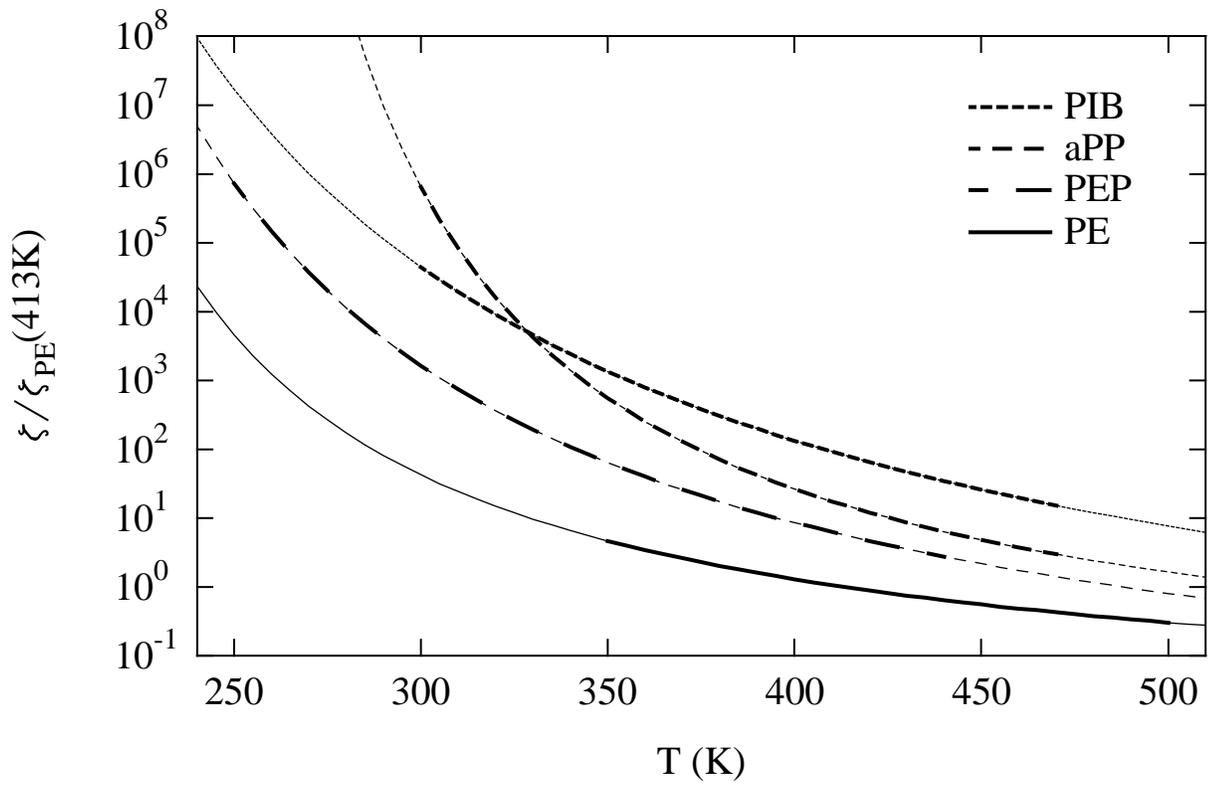,width=16cm}
\vspace{1cm}
\caption{Temperature dependence of the friction coefficients 
$\zeta(T)/\zeta_{\rm PE}(413~{\rm K})$ extracted from experimental data.
The heavy lines in the graph indicate the temperature range over which
experiments were performed.
cf.\ Table~\protect\ref{tab2}.}
\label{fig6}
\end{figure}

The temperature dependent friction coefficients are now obtained from
\begin{equation}\label{exp5}
\log(\zeta(T))=\log(\zeta(T_{\text{ref}}))
-\frac{C_1(T-T_{\text{ref}})}{C_2+T-T_{\text{ref}}}
\end{equation}
with the $\zeta(T_{\text{ref}})$ values presented in Table~\ref{tab2}.
In Fig.~\ref{fig6} we present the friction coefficients $\zeta(T)$ 
divided by the reference value $\zeta_{\rm PE}(413~K)$ for the polyolefins
considered in this work.
The heavy lines in the graph indicate the temperature range in which
experiments were performed and where the temperature correlations are 
expected to be most reliable.
For each of the polyolefins, a strong increase in the 
friction coefficient is evident as the temperature is lowered. 
This increase is due to the slowing of the dynamics of the polymers 
as the glass transition is approached. 
The glass transition temperature $T_{\rm g}$ of polypropylene (aPP) is much
higher (cf.\ Table~\ref{tab2}) than that of the other three polyolefins
considered here. This is apparent in Fig.~\ref{fig6}, where the 
friction coefficients of PE, PEP, and PIB have very similar temperature
dependencies, while the aPP friction coefficient curve starts turning
up at a much higher temperature and crosses the curve of the friction 
coefficient for PIB.  

\begin{figure}
\psfig{figure=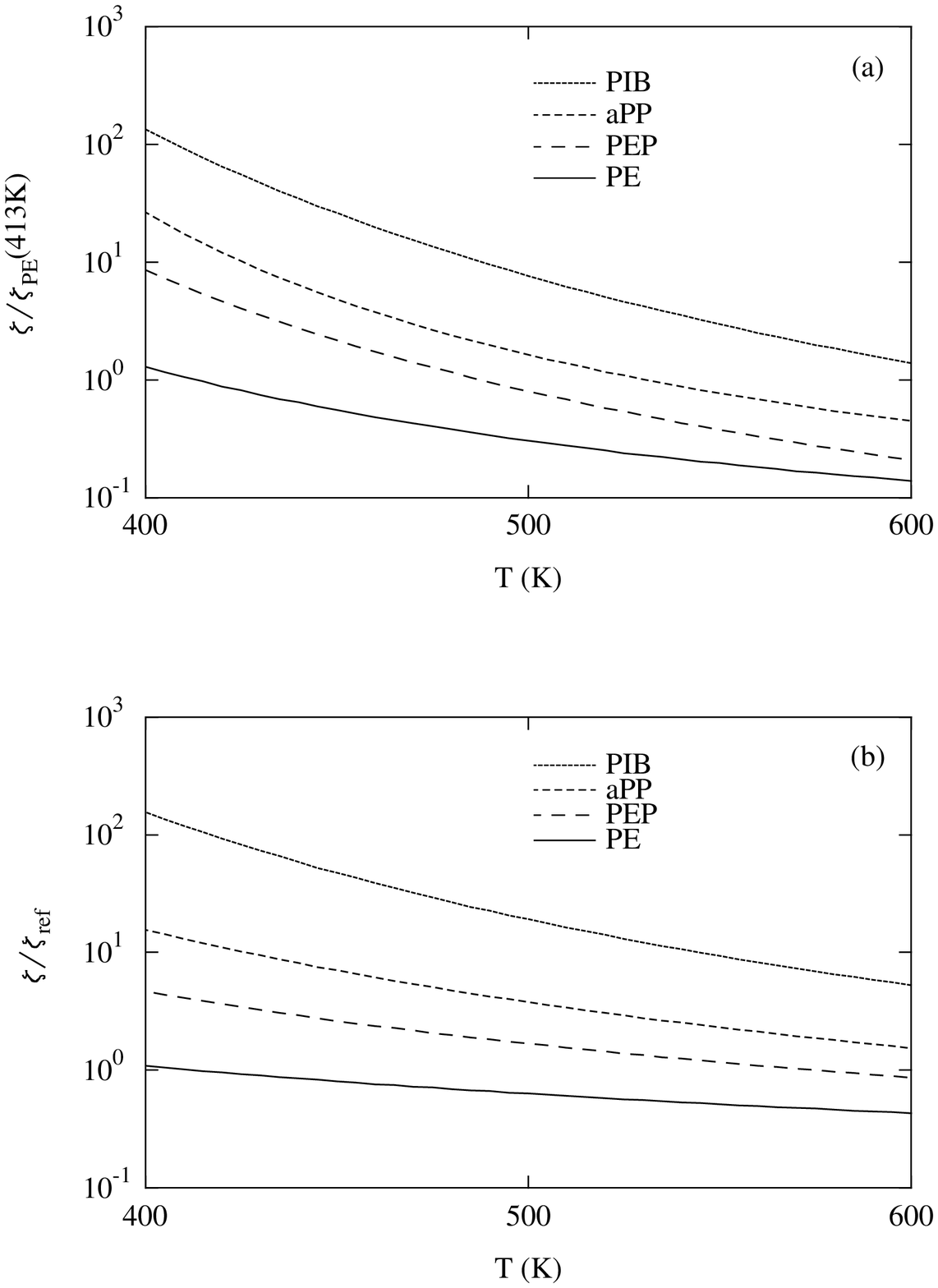,width=15cm}
\vspace{1cm}
\caption{Friction coefficients extracted from experimental
data (a) and predicted from our simulation procedure (b) as a function
of temperature at atmospheric pressure.
}
\label{fig7}
\end{figure}

In Fig.~\ref{fig7} we present the friction coefficients extracted
from experimental viscosity data together with those predicted 
from our simulation procedure for temperatures well above the 
glass transition temperatures of the polymers. Comparing the predicted
with the ``experimental'' curves we note that in both graphs the values
of the  friction coefficients increase for a given temperature 
when going from PE, which has the lowest friction
coefficient, over PEP, PP, to PIB. As noted earlier, this can be understood 
as a result of the 
different small scale architectures since the number of side groups
in the repeat unit increases from PE (0) over PEP (1), PP (2) to PIB (4).
The magnitude of the architecture effect is similar in the predicted 
and experimental friction coefficients. Furthermore, we note that the
variation of the friction coefficients with temperature is of the same
order of magnitude in the predicted and extracted curves.
While the temperature range in Fig.~\ref{fig7} is well above the glass
transition temperatures, lower temperatures 
are included in Figs.~\ref{fig4} and \ref{fig6}. As can be seen from
these figures, the agreement between 
the simulation results and the $\zeta$ values extracted from experimental
data diminishes as the glass transition temperature is approached. 
This is because
our simple approach focuses on individual segmental motion rather than
cooperative effects, a point which will be discussed below. 

\begin{figure}
\psfig{figure=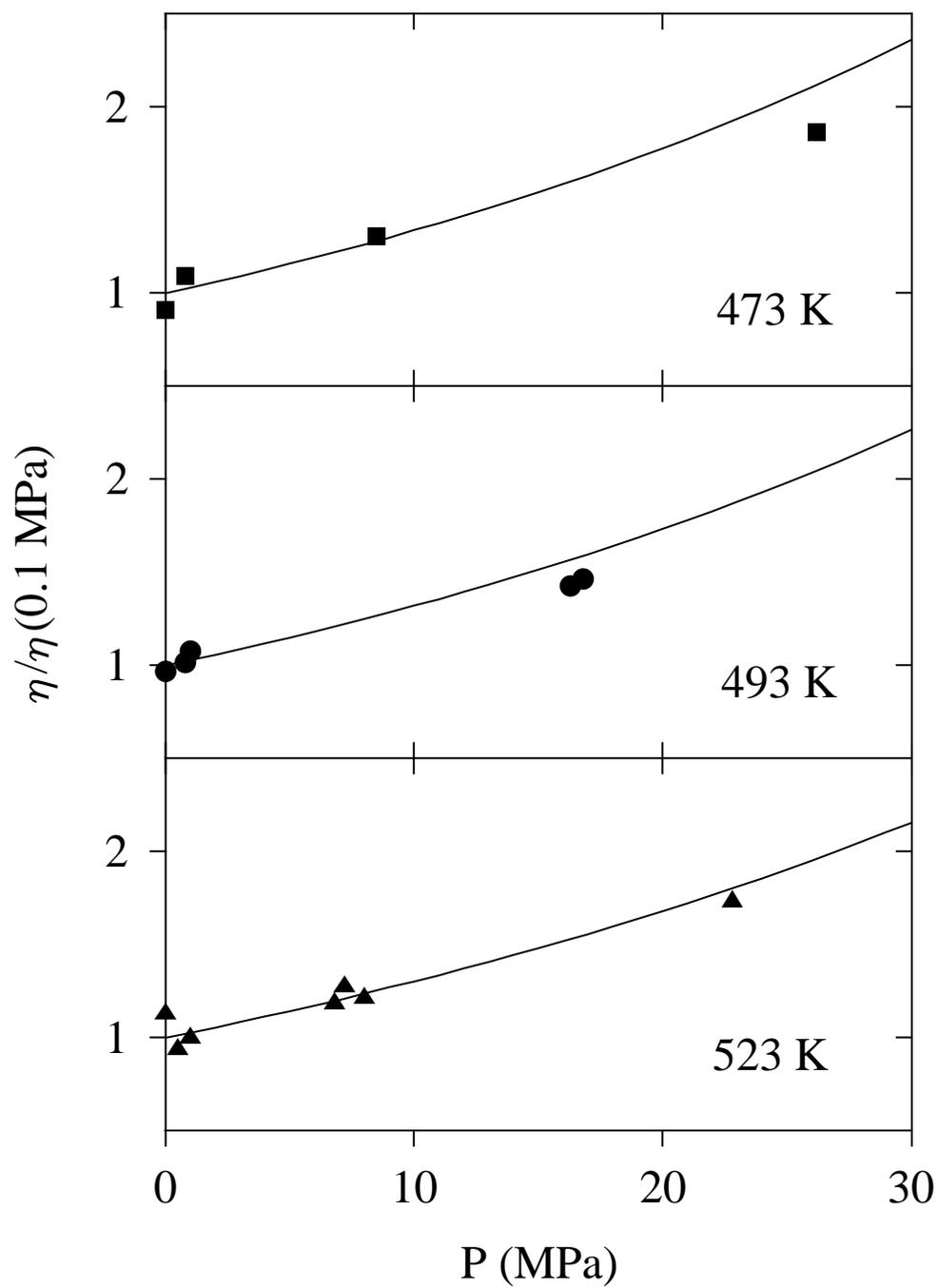,width=13cm}
\vspace{1cm}
\caption{Pressure dependence of the viscosity of polypropylene.
For each of the three isotherms we present viscosities 
divided by their value at atmospheric pressure.
The lines represent results from our simulation procedure,
the symbols indicate experimental data by Mattischek and 
Sobczak.\protect{\cite{ma97e}}}
\label{fig8}
\end{figure}

In Fig.~\ref{fig8} we compare the pressure dependence of the viscosity 
of polypropylene 
as predicted from our work with experimental data~\cite{ma97e} 
along three isotherms.
To separate temperature and pressure effects, we use the values of the
viscosity at atmospheric pressure to scale the viscosities on each 
isotherm. For low to moderate pressures ($\leq$20~MPa) the agreement is
excellent. As the pressure increases further, our work overestimates the 
viscosity. This is likely due to our employing the BGY-lattice-model 
equation of state with parameters optimized for low pressures,\cite{lu99}
an issue which will be addressed in future work.

\section{Discussion}\label{disc}

In this work, we presented an exact enumeration method for 
lattice simulations of chain segment mobility.
The algorithm enumerates the attempted and successful 
moves for two short, straight sections of a polymer and is evaluated 
by taking relative frequency, energetics, and density effects 
into account. The result is the mean probability for 
segmental motion as a function of reduced temperature and
the filling fraction of the lattice. 
We performed simulations for four different small-scale architectures 
obtaining results that show a sensible decrease in mobility
with increasing density and number of side groups of a monomer.

Combining these results with equations of state for the corresponding
polyolefins, we deduce monomeric friction coefficients as a function 
of temperature and pressure. Our method is not an absolute one, but
employing the friction coefficient of polyethylene at 413~K and atmospheric
pressure as a reference value, we can predict the relative values of the 
friction coefficients as a function of temperature and pressure for 
the polyolefins considered in this work.  
If we are interested in the properties of a single polyolefin, we can
employ the value of the viscosity for a particular temperature
and pressure as a reference value and predict the relative variation of 
the viscosity with temperature and pressure from there. 
The same is true for other transport properties that depend in a
simple way on the friction coefficient; this will allow us to investigate
diffusion coefficients, for example, in future work.

To compare our results with experimental data, we extracted Rouse 
viscosities and monomeric friction coefficients from high molecular 
mass viscosity data. Employing temperature correlations of the experimental
data in the WLF form we obtain ``experimental'' monomeric friction coefficients
at atmosperic pressure over a range of temperatures, which we scale
by the value for PE at 413~K. 
The comparison of these extracted friction coefficients with the 
results from our new simulation method is encouraging:
For temperatures well above the glass transition temperature
the calculated probabilities give a good qualitative
representation of the relative variation of the friction coefficient
with temperature and monomer architecture.
A comparison of calculated and experimental pressure variation of the 
viscosity of polypropylene along three isochores shows excellent results
for lower pressures.

In order to extend the range of validity of the present theory to 
temperatures closer to the glass transition, 
cooperative effects in the dynamics will have to 
be taken into account in a more sophisticated way.  In this first work
with the new simulation method, we have assumed a random distribution of voids
over the lattice and take the system to be in an equilibrium state
before each attempted move. One way to improve on this approximation 
would be to
employ an iterative approach in which the distribution of configurations
after a round of attempted moves is used as the input distribution 
of configurations for the next round of moves. 
Finally, the simulation method introduced here was applied only 
to straight chain sections of 
polymers on a cubic lattice with a single site-site interaction 
strength $\epsilon$. It is, however, readily modified to include
chain flexibility, chemical differences and realistic bond angles,
which allows a large range of polymeric systems to be investigated in
this way. 
We plan to extend the theory in these directions and are currently 
focusing on the effects of chain flexibility.

\section*{Acknowledgements}
Financial support through a faculty research grant from the University of
Akron is gratefully acknowledged.

\begin{table}
\caption{BGY lattice-model parameters for the polyolefins considered in
this work.\cite{lu99} Please note: 
The value $r$ for the number of lattice sites 
per chain is proportional to the molar mass of the polymer. The values 
presented here correspond to a molar mass of $M=170,000$ for each of
the polyolefins.}
\begin{tabular}{|c|c|c|c|c|} 
Polymer & PE & PEP & aPP & PIB \\ \hline
$\epsilon$(J/mol)  & -1977.5 & -2000.0 & -2040.7 & -2208.1 \\
$r$ & 14210.6 & 14226.1 & 14471.8 & 13433.8 \\
$v$(L/mol) & 0.013 & 0.013 & 0.013 & 0.013
\end{tabular}
\label{tab1}
\end{table}

\begin{table}
\caption{Experimental parameters for the polyolefins considered in
this work. The values correspond to atmospheric pressure
and a temperature of 413.15~K. The glass transition temperature
for polyethylene (PE) is an estimate based on results for ethylene-butene
copolymers.\cite{ca84} }
\begin{tabular}{|c|c|c|c|c|} 
Polymer & PE & PEP & aPP & PIB \\ \hline
$R^2_0/M$ (\AA$^2$mol/g) (Ref.~\onlinecite{fe94})
&  1.21 & 0.834 & 0.67 & 0.57 \\
$\rho$ (g/cm$^3$) (Ref.~\onlinecite{fe94})
& 0.785 & 0.79 & 0.791 & 0.849 \\ 
$M_{\rm e}$ (Ref.~\onlinecite{fe94}) & 976 & 2284 & 4623 & 7288 \\ \hline 
viscosity data references & ~\onlinecite{pe87,pe94} & 
~\onlinecite{ge97} & ~\onlinecite{pe88} & ~\onlinecite{fe91} \\
temperature range T(K)  & 
350--500 & 248--443 & 298--463 & 298--473  \\ \hline
$\log(\frac{\displaystyle \eta_{\rm R}}{\displaystyle M}/
\frac{\displaystyle \rm{Poise}}{\displaystyle \text{g/mol}})$  
& -4.1035 & -3.6039 & -3.3424 & -2.7669 \\ 
$\log(\zeta/\text{Poise cm})$ 
& -8.5554 & -7.8001 & -7.3648 & -6.6388 \\  \hline 
$C_1$ & 2.018 & 3.565 & 3.101 & 4.684  \\
$C_2$ (K) & 253 & 277 &  189 & 307 \\ \hline
$T_{\rm g}$ (K) & 188 & 211 & 268 & 202   \\
$T_{\rm g}$ references & ~\onlinecite{ca84} & ~\onlinecite{ge97} &
~\onlinecite{pe88} & ~\onlinecite{be68,ng96}
\end{tabular}
\label{tab2}
\end{table}

\end{document}